\documentclass[journal,twocolumn]{IEEEtran}
\usepackage{cite}
\usepackage[cmex10]{amsmath} 
\usepackage{color}
\ifCLASSINFOpdf
  \usepackage[pdftex]{graphicx}
\else
  \usepackage[dvips]{graphicx}
\fi
\usepackage{fixltx2e}
\usepackage[tight,footnotesize]{subfigure}
\usepackage{url}
\usepackage{verbatim}

\begin{document}

\title{Low Complexity Indoor Localization in Wireless Sensor Networks by UWB and Inertial Data Fusion}

\author{\IEEEauthorblockN{Alberto Savioli, Emanuele Goldoni, Pietro Savazzi, and Paolo Gamba}\\
\IEEEauthorblockA{University of Pavia\\
Dipartimento di Ing. Industriale e dell'Informazione\\
Via Ferrata 1 - 27100 Pavia, Italy\\
Email: \{name.surname\}@unipv.it}}

\maketitle

\begin{abstract}
Precise indoor localization of moving targets is a challenging activity which cannot be easily accomplished without combining different sources of information. In this sense, the combination of different data sources with an appropriate filter might improve both positioning and tracking performance.
This work proposes an algorithm for hybrid positioning in Wireless Sensor Networks based on data fusion of UWB and inertial information. A constant-gain Steady State Kalman Filter is used to bound the complexity of the system, simplifying its implementation on a typical low-power WSN node. 
The performance of the presented data fusion algorithm has been evaluated in a realistic scenario using both simulations and realistic datasets. The obtained results prove the validity of this approach, which efficiently fuses different positioning data sources, reducing the localization error. 
\end{abstract}

\section{Introduction}\label{sec:intro}
Wireless Sensor Networks (WSNs) localization applications are usually related to the need of tracking the positions of some target nodes moving into a delimited area, such as an industrial warehouse, a museum or a battlefield.
These applications may need different accuracies for the localization process: since tracking a visitor in a museum is not a critical issue, localization exploiting radio signals may be acceptable, on the contrary cart driving needs precise information about spatial coordinates and dynamics. In the latter case, data coming from a single source might not be accurate enough, and sensor fusion can be exploited to achieve higher precision \cite{WPRB, Kianoush2012}.

Usually, fusion starts from measurements coming from different sources, such as radio, inertial, ultra-sonic, GPS, and optics, although most of them are not suitable for a wireless sensor device due to cost limits and energy budget. Hence, a suitable way to operate in a WSN with a mobile target is to fuse radio measurements with inertial ones. If the additional hardware needed for the inertial systems is based on Micro Electro-Mechanical Systems (MEMS) technology, costs and complexity are still affordable. 

For example, the authors of \cite{Herrera2007} propose the adoption of a Kalman filter for combining time of arrival (TOA) Ultra-Wideband measurements with a low-cost MEMS inertial unit. While in \cite{Hol2009, Pittet2008} the Kalman filter is replaced by an Extended Kalman filter (EKF), which provides better results for nonlinear sistems. Similarly, an EKF is used in \cite{Corrales2008} to fuse an inertial motion capture system with UWB measurements. In more details, the UWB system drops TOA and uses a combination of time-difference of arrival and angle of arrival (AOA) techniques to estimate the 3-dimensional global position of the user.

In this paper we propose a data fusion algorithm able to combine data coming from different sources, namely UWB TOA measurements and inertial data. Differently from previous works, the presented approach is based on a Steady State KF with a fixed gain, which can be easily implemented and run on typical low-power and low-performance WSN nodes.

The algorithm has been evaluated in a realistic scenario, evaluating its performances both through simulations and using measurements collected on the field during a measurement campaign.
Results confirm that fusing different sources of information for localization purposes is desirable, and the proposed approach can provide notable advantages without significantly increasing the computation overhead.

The remainder of the article is structured as in the following. In Sections \ref{sec:radio} and \ref{sec:imu} we briefly overview localization algorithms using data fusing. Then, Section \ref{sec:fusion} describes in details the proposed algorithm for fusing positioning information coming from inertial and radio sensors. Experimental performance evaluation is presented in Section \ref{sec:experimental}, while Section \ref{sec:conclusion} includes some final remarks.

\section{Radio Localization and Tracking}\label{sec:radio}
Localization in WSNs usually refers to the determination of the unknown position of a target node using radio information. The most common approach involves a limited number of nodes, called anchors, that are aware of their positions -- the unknown device can be localized estimating the distances between the target and the anchors. 

Radios might be employed for localization in different ways, depending on the network communication technology. For instance, the widely-used IEEE 802.15.4 narrowband standard allows distance estimation by exploiting the received signal power strength, the known transmitter power, and the path loss model.
Further, the larger bandwidth available for the Ultra Wide Band IEEE 802.15.4a radios can be used to achieve a higher spatial resolution, by sending very short pulses.

Range measurements among nodes are used to find their positions in the deployed area. This is done through a localization algorithm, whose input is a pool of distance measurements referred to the anchors, and its output is the estimated position of the unknown node.
A number of localization algorithms can be found in literature: some of them are based on simple geometric considerations, while more complex ones exploit statistical methods that make the algorithm more robust to noise at the expense of an increased computational complexity. Therefore, a trade-off between performance and complexity should be taken into account for realistic applications.
For example, multilateration \cite{Zanca2008} is a geometry-based algorithm. The estimated distances $d_i$ are used to create circles centered in the anchor nodes with a radius $r$ equal to $d$. Ideally, the target node should be exactly placed at the intersection of all the circumferences.
The Min-Max algorithm \cite{Lymberopoulos2006, Langendoen2003} is also based on simple geometric considerations, and it is equally widely used for localization due to its easiness of implementation. In this algorithm, the estimated distance $d_i$ is used to define squares as opposed to circles in a multilateration algorithm. 

The noisy channel measurements degrade the localization accuracy, so tools like Bayesian or Particle filters might be employed to increase the overall performance.

Bayesian filters are based on the recursive estimation of the system status, corresponding to the  target node position, while the status observations are the sensor measurements \cite{Fox2003}. In these filters, the status update is performed each time a new observation is available: once a new measurement comes, the filter updates the system status with this new information, taking into account its history. Further, this types of filters are perfect tools for a multi-sensorial localization system, since they are capable of working over different sources

However, this family of filters can converge to the true posterior distribution only if considering linear systems and gaussian noise. These constrains represent a big limitations for real applications. Hence, particle filters (PFs) represent an implementation of recursive Bayesian filters that can be also used for non-linear and non-gaussian problems \cite{Fox2003, Gustafsson2002}. In these systems, the posterior distribution density is represented through samples -- the particles -- having different weights.

Since PFs accuracy is strictly related to the number of particles, this technique is usually implemented when memory is not a main issue and hundreds of particles can be used. Some researchers have investigated how to modify these filters in order to reduce the required memory footprint. For example, S. Mazuelas et al. in \cite{Mazuelas2011} proposed the Belief Condensation Filter (BCF) that uses gaussian distributions as particles. They proved that using a condensation of gaussian distributions it is possible to obtain good accuracies using less particles than with a standard PF approach, thus saving memory.

\section{Inertial tracking with IMU}\label{sec:imu}
Micro-Electro-Mechanical Systems (MEMS) sensors are used nowadays in a wide range of applications, from robotics to entertainment. Although less accurate than traditional sensors, MEMS are chosen because they are lighter, cheaper and more robust to stresses. Moreover, they require a small amount of energy to work and they are precise enough for many WSNs applications. 

Recently, MEMS technology has also been adopted to equip Inertial Measurement Units, replacing traditional electro-mechanical sensors. An Inertial Measurement Unit (IMU) is a device able to measure velocity, orientation, and gravitational forces of a body, using a combination of accelerometers gyroscopes, and sometimes also magnetometers. Then, using a method known as dead reckoning, data collected from IMU's sensors allow to track a craft's position.

\subsection{Errors in MEMS IMUs}
With the raw data coming from these sensors, it is possible to reconstruct the movement of a body. Specifically, accelerations have to be integrated twice to obtain the covered space, while a single integration is needed to get the rotation angle from the revolution speed sensed by the gyroscopes. However, MEMS sensors are affected by noise, and the long-term effect of their inaccuracy is not negligible.

The operating principle behind a MEMS accelerometer is the displacement movement of a small mass. Often, two surfaces are used as a variable capacitor, and the output voltage depends on the distance between two planar surfaces. The outcomes of these sensors are affected by ripples around the actual values, resulting in an growing error due to the double integration needed to transform accelerations into displacements. Given an average noise $N$, the space estimation error $e$ will rise over time $t$ according to
\begin{equation}\label{eq:accel-error}
e(t)=N \cdot \frac{t^{2}}{2}
\end{equation}
Hence, periodic recalibration of the position using GPS or similar technologies is needed.

Gyroscopes measure the rotation impressed on a body, by reading directly the angles or the angular speed.
MEMS gyros are built with a variable capacitor, whose moving plates are composed by a silicon cantilever connected to a vibrating mass. If a rotation with angular speed $\omega$ is impressed to the body, the mass is subjected to the Coriolis force $F_c$:
\begin{equation}\label{eq:coriolis}
F_{c}=-2m(\omega \times v)
\end{equation}
where $v$ is the average speed of the vibrating mass and $m$ its weight. This force makes the mobile plate of the capacitor move, and the capacitance changes accordingly.
This type of gyroscopes output a voltage proportional to the angular speed, and an integration is needed to obtain the revolution angle.
The main error source of these sensors is given by the \emph{bias}, that is the output value that it gives even if no rotations are impressed to the device. This source of error $\nu$ generates a drift of the angle $\theta(t)=\nu \cdot t$ that grows linearly over time, leading to misestimations of the angle. This error can be modeled by recording the output of the sensor in static conditions, and extracting the constant component with ML fitting. 

Since gyroscopes are affected by biases, IMUs are often equipped with a magnetometer that can be used to recalibrate the gyroscopes by zeroing the integral term when the body is headed in a known position. 
The sensor output is a vector that points the magnetic North, so it can be exploited to bound the cumulative errors of gyroscopes by comparing the estimated heading of the device from the North with the one given the magnetometer. The difference in this angle is the amount of drift accumulated by the gyro in the considered amount of time.
The working principle is based on the movement of a micro-structure, due to the Lorentz force acting on a micro-conductor that carries the current in a magnetic field. Electric and temperature noise affect MEMS magnetometer, but no integration is needed since these sensors are not affected by cumulative errors.

\subsection{Dead Reckoning}
\begin{figure}[tbp]
	\centering
	\includegraphics[width=0.35\textwidth]{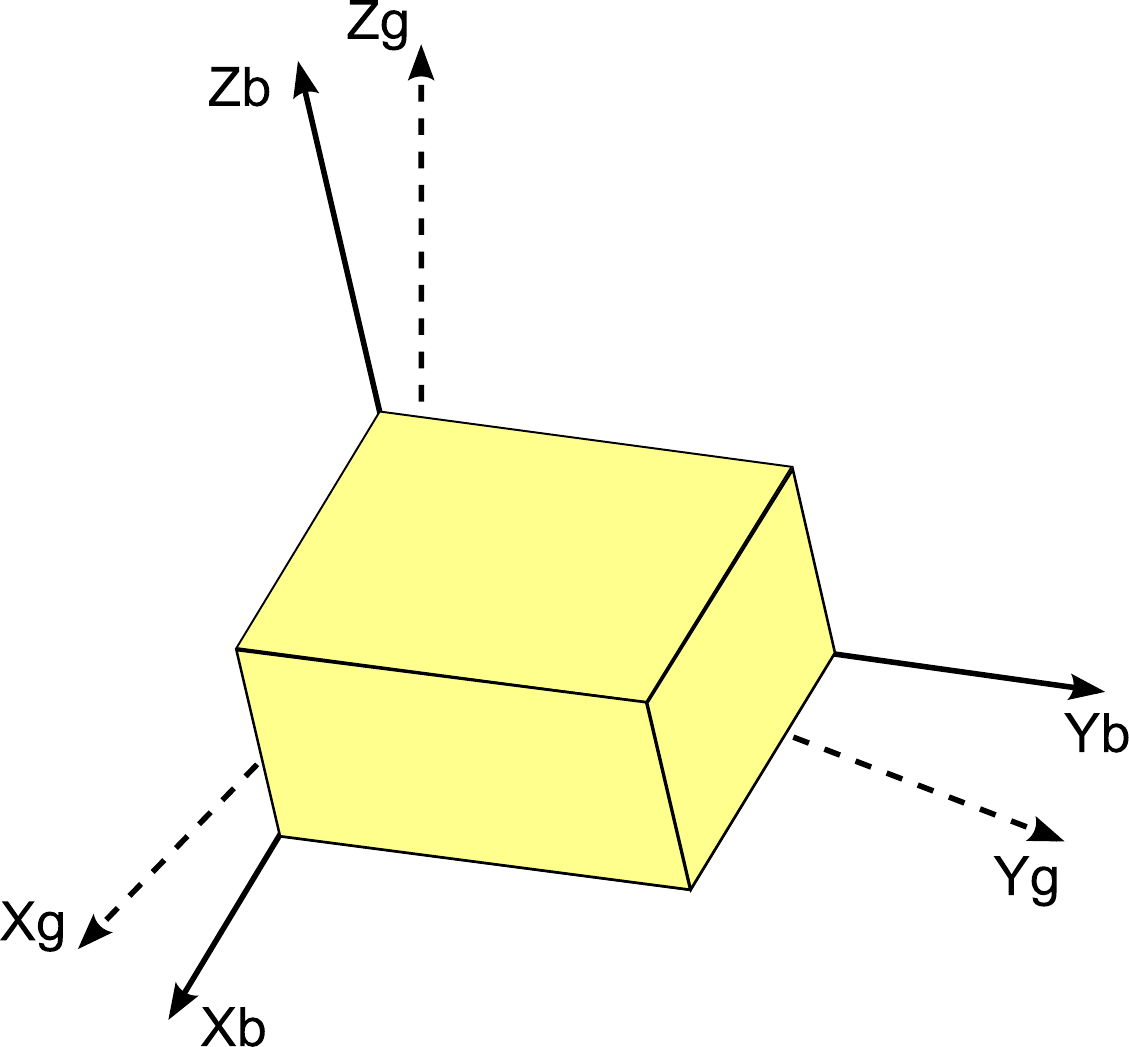}
	\caption{\label{fig:sisrif}Global and Body reference systems.}
\end{figure}

Dead reckoning is exploited to estimate the state of a mobile device by using the previous inertial values. Usually, the state is described by the inertial measurements coming from the IMU, such as accelerations and heading, for each of the three space axis. If a device traveling in a 2D environment is considered, only one rotational information is needed to estimate the heading.
The state vector $X$ and the measurement vector $U$ of a mobile device may be described by:
\begin{equation}\label{eq:vec}
\begin{array}{lr}
X' = \begin{bmatrix}\ddot x, \dot x, x, \ddot y, \dot y, y, \ddot z, \dot z, z, \dot \theta, \theta\end{bmatrix}\\
U' = \begin{bmatrix}1, 0, 0, 1, 0, 0, 1, 0, 0, 1, 0\end{bmatrix}
\end{array}
\end{equation}
where $\ddot x, \ddot y, \ddot z$ represent the linear inertial measurements along the three axes and $\theta$ the rotation of the device over the $z$ axis. From the measurement vector $U$ it can be evinced that the measured entities are the accelerations and the angular velocity coming from the IMU. The other elements of the state are computed using Newton's motion laws, deriving speeds and distances from the accelerations, and angles from the angular speeds:
\begin{equation}\label{eq:newton}
\begin{array}{l}
x_{k+1} = x_0 + \dot x_k \cdot t + \frac{1}{2} \ddot x_k \cdot t^2 \\
y_{k+1} = y_0 + \dot y_k \cdot t + \frac{1}{2} \ddot y_k \cdot t^2 \\
z_{k+1} = z_0 + \dot z_k \cdot t + \frac{1}{2} \ddot z_k \cdot t^2 \\
\theta_{k+1} = \theta_0 + \dot \theta_k \cdot t
\end{array}
\end{equation}
Usually, IMUs are fixed on a device so that their axes overlap. By this way, all the measurements coming from the sensors are referred to the body coordinate system, that is different from the global one. Specifically, two coordinates systems are generally used, one referred to the body frame $(X, Y, Z)_b$ and the other to the global frame $(X, Y, Z)_g$. Fig. \ref{fig:sisrif} shows the configuration of the two systems.

In order to reconstruct the device movement in the global frame, the acceleration vectors sensed by the IMU in body coordinates have to be referred to the global frame. To do this, a Rotation Matrix that relates the two systems is used. This matrix is a combination of rotations that allows to project a vector in the body frame to the one in the global frame. An example of vector projection from the body to the global frame, for rotation around the $Z$ axis is given by:
\begin{equation}\label{eq:rot}
u_g = R \cdot u_b
\end{equation}
where $u_g$ and $u_b$ are the vectors referred to the global and the body frame, respectively, and $R$ is the rotation matrix for an angle $\Phi$, defined as:
\begin{equation}\label{eq:rot1}
R = \left[ \begin{array}{ccc}
					0 & \textrm{cos}(\Phi) &  -\textrm{sin}(\Phi)\\
					0 & \textrm{sin}(\Phi)  & \textrm{cos}(\Phi)\\
					1 & 0 & 0 \\
					\end{array}
					\right]
\end{equation}

\section{Radio and Inertial Sensors' Fusion}\label{sec:fusion}
As described in Section \ref{sec:radio}, commonly-employed strategies for fusing the inertial measurements with those coming from UWB radios are, for example, particle filters or Bayesian estimators. The main disadvantages of these techniques are the high amount of memory required to store the probabilistic distributions of each particle, and the high complexity caused by large matrix manipulations. To cope with these issues, we advocate the use of a steady state Kalman Filter (KF). This choice limits the complexity of the tracking algorithms, allowing to implement and run it over a low-power embedded processor.

The fusion algorithm proposed in this work is based on three main steps. Firstly, we compute the displacement at each point in global coordinates using inertial measurements. Then, localization is performed using the corresponding set of UWB TOA range measurements and the Min-Max localization algorithm. Finally, the proposed steady state KF is used to correct the positions obtained by the inertial ones using UWB measurements. Since our target is a wheel cart moving on a floor, the z-axis can be safely ignored. However, the rest of the analysis could be easily extended to a three-dimensional movement including, if needed, a third axes.

The first step of the process estimates the distance covered by the mobile device at each sampling time, with its direction. This is actually the dead reckoning phase, where the target motion is computed starting from the dynamics sensed by the inertial sensors. Hence, starting from the previous estimation $(x_k, y_k)$, a new pair of coordinates $(x^{imu}_{k+1}, y^{imu}_{k+1})$ are produced accordingly to the sensed inertial data. This can be easily done by using accelerations, and then computing the velocity and the space covered by the target node using Eq. \eqref{eq:newton}, additionally considering the information about angular velocity.

In the second step, the position of the target device is estimated by using the range measurements coming from the radio receivers. Specifically, TOA measurements using UWB devices are considered, and the Min-Max algorithm is used to join the ranges referred to the same instant in order to estimate  the coordinates $(x^{UWB}_{k+1}, y^{UWB}_{k+1})$ of the new position of the target node.

Finally, a steady state Kalman Filter is used. As already mentioned, a simplified version of KF is proposed to keep the computational complexity as lower as possible. 

The filter state vector at time $k$ is given by the coordinates of the target device:
\begin{equation}\label{eq:stateKF}
X_k=
	\begin{bmatrix}
		x^{target}_k \\ 
		y^{target}_k
	\end{bmatrix}
\end{equation}
Since all the calculations about the system dynamics have been already performed, the KF will simply decrease the covariance between the sets of measurements coming from the IMU and radio, by filtering the measurement noise. 
Using the simplified version of the filter, only 2 by 2 matrices are employed instead of larger ones needed if the system dynamics would be computed by the KF. Specifically, the dynamics matrix $H$ of this filter is represented by the unitary 2x2 matrix: 
\begin{equation}\label{eq:HKF}
H_k=\mathbf{I}_k=
	\begin{bmatrix}
		1 & 0 \\ 
		0 & 1
	\end{bmatrix}
\end{equation}
The measurements vector $U_k$ is populated by both inertial and UWB measurements: the filter will use the new measurements when available, to update the state estimation and compute the new position.

Since the implemented filter is a steady state KF, the gain is computed off-line and kept constant. The fixed Kalman gain $K$, referred to the system dynamics, can be saved as a constant two-elements vector.

\section{System Evaluation}\label{sec:experimental}
Firstly, a realistic IEEE 802.15.4a channel and the readings coming from an inertial platform have been simulated, to evaluate the presented approach validity. Then, in order to consider a real indoor scenario, we have applied the novel algorithm to the measurements collected  using the network set up in 2009 for the Newcom++ project \cite{WPRB}.

\subsection{Scenario}
\begin{figure}[htb]
		\includegraphics[width=3.4in]{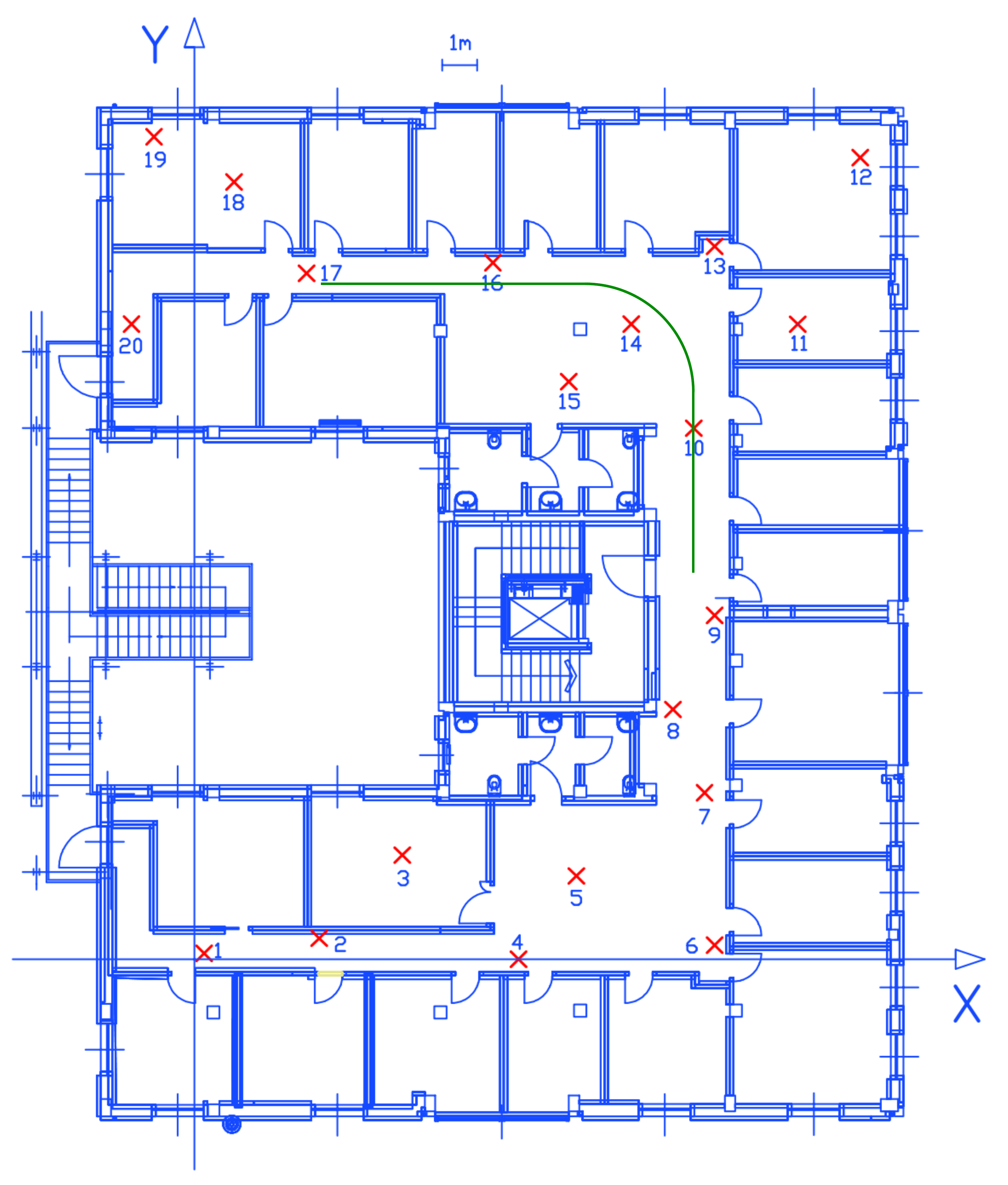}
        \caption{Location of UWB anchors (red markers) and path of the moving target (green curve) \cite{WPRB}.}\label{fig:floor}
\end{figure}

The collection of real measurements was carried out in 2009 within the Newcom++ WPR.B workpackage. The network was installed inside a building of the University of Cesena, Italy, and it included 10 Ultra Wide Band nodes, one inertial measurement unit, and some ZigBee nodes. The UWB anchors were scattered on the floor: some of them were in line of sight (LOS) along to path followed by the moving device. The remaining anchors were instead deployed inside rooms, providing possible non LOS (NLOS) measurements. The mobile device -- a 4 wheels cart following a prefixed path of about 18 meters -- was equipped with a wireless radio-frequency module and IMU. The robot movement along the path was recorded, and the radio signals coming from the anchors and the inertial data was also collected. A floor plan of the measurement scenario is shown in Figure \ref{fig:floor}.

Among all the possible testbed configurations, the most interesting one refers to a dynamic scenario where the target device is continuously moving along the path with an average speed of 0.29 m/s, with a variable instantaneous velocity ranging from 0.1 m/s to 0.5 m/s. 
TOA measurements coming from UWB radios were recorded at a sampling frequency of 2 Hz. The original sampling rate for the inertial unit was 500Hz. Since this rate is not applicable to real-world WSNs, these data have been sub-sampled to 5 Hz before feeding them to the filter.

\subsection{Numerical simulation}\label{numerical}
In order to create a realistic simulation environment, the channel model of IEEE 802.15.4a standard has been used \cite{IEEEmodel}. According to experimental measurements, we assume an ideal time of arrival (TOA)-based ranging where the only noisy impairment is due to the channel model randomness. The indoor multipath channel is modeled as \cite{Alsindi2009}:
\begin{equation}
h_{m}(t)=\sum_{k=1}^{L}\alpha_k e^{j\phi_k}\delta(t-t_k),
\end{equation}
where $\alpha_k$ is the magnitude, $\phi_k$ the phase and $t_k$ the delay of the $k^{th}$ multipath component. In case of LOS channel, the direct path between the anchor node and the mobile terminal simply corresponds to $\alpha_0, t_0$, while for NLOS channels  we consider $\alpha_j, t_j$, corresponding to to strongest multipath component, i.e.:
\begin{equation}
\label{eq2}
\alpha_j=\max_k(|\alpha_k|). 
\end{equation}
Once the direct path delay is determined, the distance between the terminal and the $m^{th}$ anchor is simply computed as in the following:
\begin{equation}
d_m=c \times t_m,
\end{equation}
where $t_m$ is set to $t_0$ or $t_j$ according to respectively a NLOS or LOS channel.
Actually, the number of multipath components in indoor channels is very high. In order to consider possible ranging errors also for the LOS case, we used Eq. \eqref{eq2} for both the channel conditions, considering the power multipath component spread around its maximum.

We have also simulated an inertial unit with accelerometers afflicted by a measurement additive gaussian noise $N$ and a sampling rate equal to 5 Hz. We simulated different values of $N$ over the whole path, and 20 simulations were performed for each setup. Comparing the output of the real accelerometer and the simulated one, we found that an SNR of $60 dB$ minimizes the difference between the two datasets. Although the obtained results are close to the experimental values, the simulation is oversimplifying -- more realistic models for IMUs could be used \cite{Quinchia2012}, but the use of a complete unit is outside the scope of this work.


Simulating a scenario similar to the considered Newcom++ deployment, a RMSE of 3.21 m has been obtained, using UWB radio localization with Min-Max. On the other hand, the average RMSE was about 1.42 m when tracking the movement of the target through the IMU.

Finally, the proposed data fusion technique is able to lower the localization root mean square error to 0.62 m. Hence, simulation results seems to prove that the steady state KF can effectively tackle noisy measurements, tracking a moving target with enough accuracy for many indoor applications.

\subsection{Experimental evaluation}

After having investigated the performance of our approach with a simulator, we have considered a real dataset contained in the WPR.B database. As in simulation, we computed the accuracy in the three cases, that is when using only Min-Max with UWB TOA ranges, when performing dead reckoning with inertial measurements, and when running our algorithm.
 
\begin{figure}[htb]
		\includegraphics[width=3.4in]{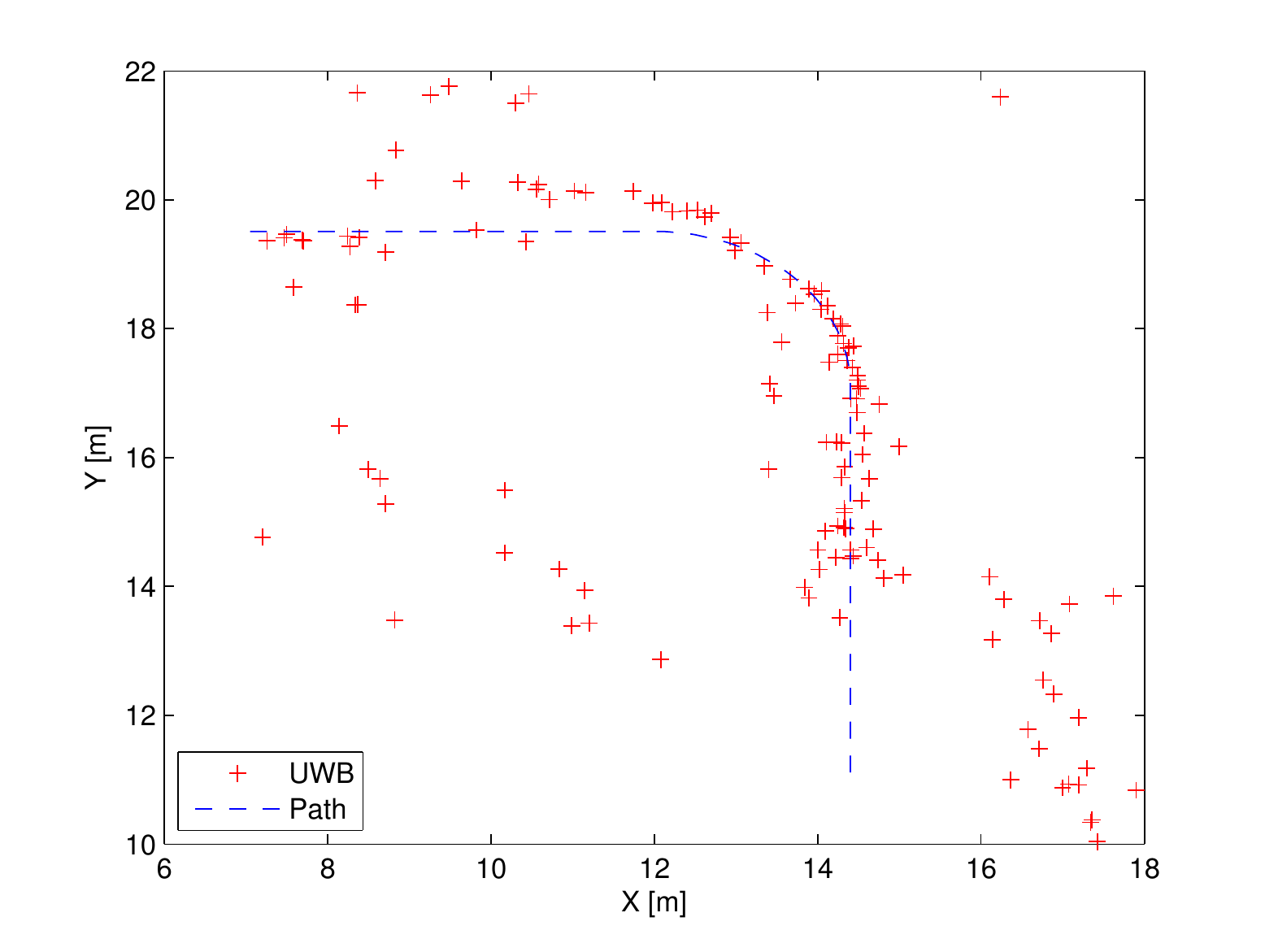} 
        \caption{Localization results and relative errors using UWB only.}\label{fig:onlyUWB}
\end{figure}

Fig. \ref{fig:onlyUWB} shows the results obtained using UWB measurements only. The position estimates are most of the time far from the right path, sometimes more than five meters far from the actual position: these values are not acceptable for indoor tracking applications. Such errors are due to the presence of walls, that cause NLOS propagation conditions between antennas, which degrade the TOA ranging accuracy. As a result, the localization RMSE provided by only Ultra Wide Band data is 3.25 m.

\begin{figure}[htb]
		\includegraphics[width=3.4in]{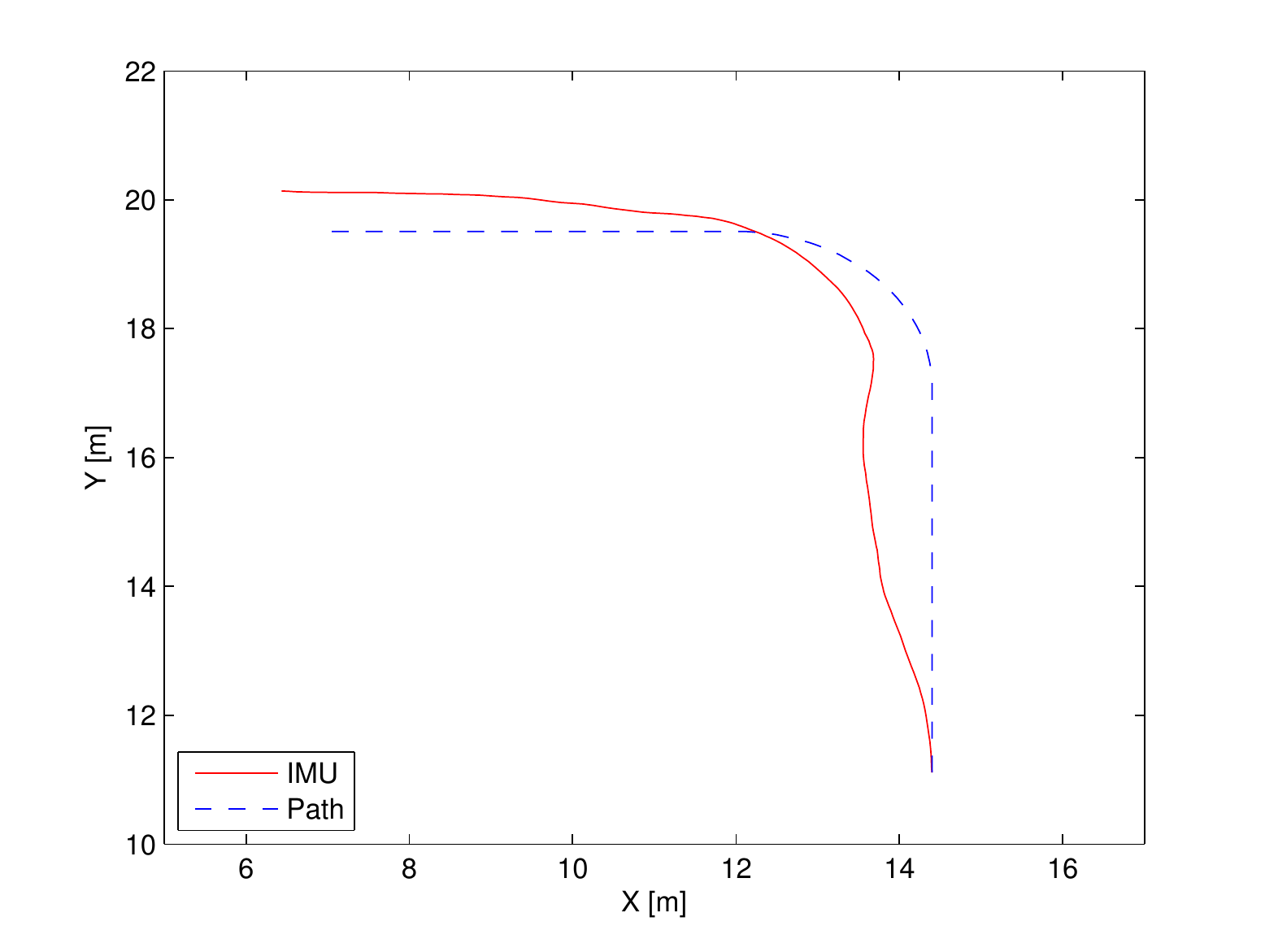} 
        \caption{Localization results and relative errors using inertial data only.}\label{fig:onlyIMU}
\end{figure}

On the other side, Fig. \ref{fig:onlyIMU} shows the results obtained by only using the inertial data. In this case, the localization process performs much better, with a maximum error bounded to 2.8 m and a RMS error of 1.22 meters. However, the drifting behavior of the inertial sensors is clearly visible, since the estimated path deviates from the real one starting from the very first part of the path. As previously noted in this Section, the cumulative error tends to grow with time: the drift effects is more evident for longest time intervals.

\begin{figure}[htb]
		\includegraphics[width=3.4in]{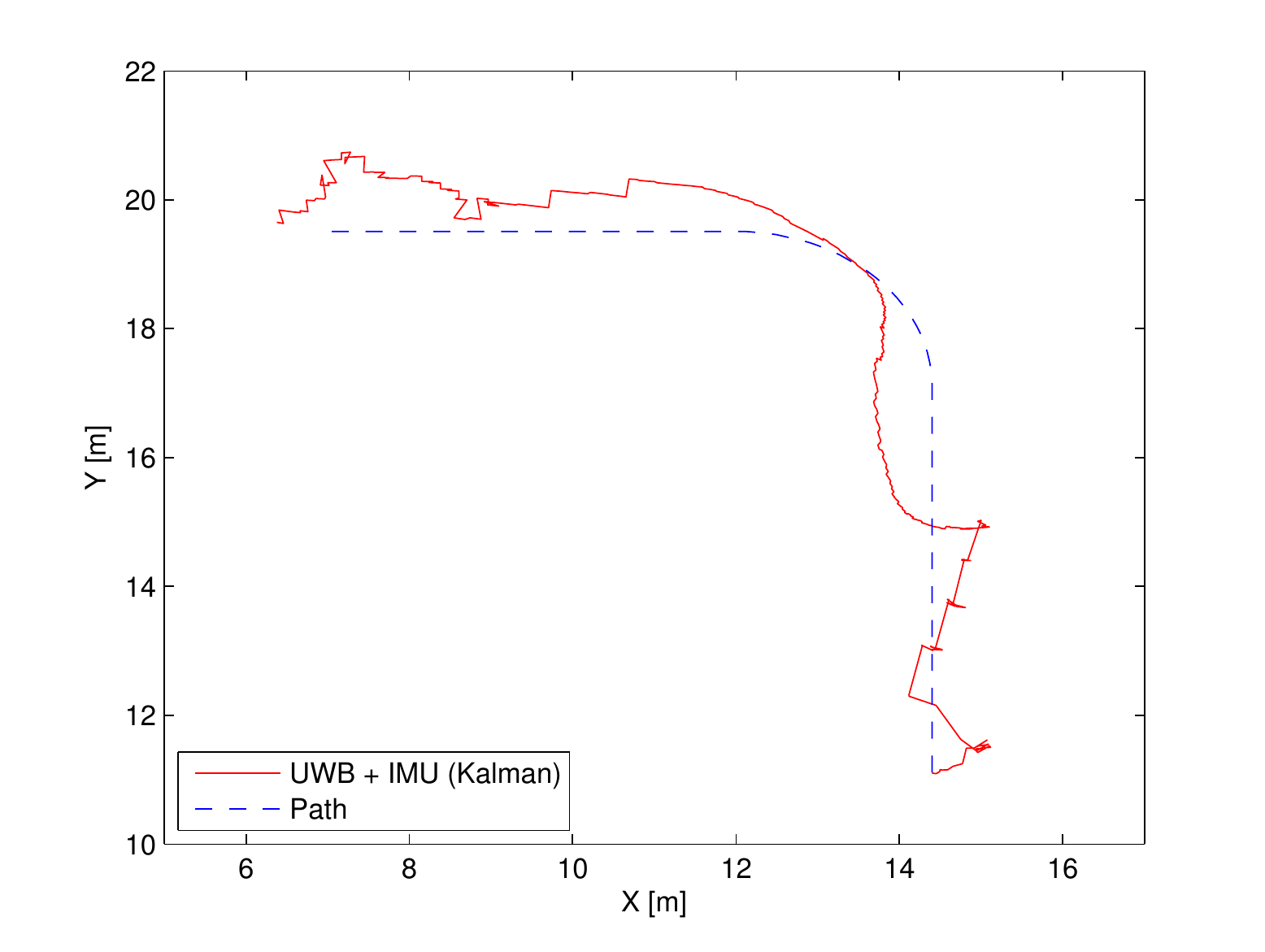} 
        \caption{Localization results and relative errors using the proposed data fusion algorithm.}\label{fig:hybrid}
\end{figure}


The tracking path output is shown in Fig. \ref{fig:hybrid}. It stands out that the tracking algorithm follows the real path during almost all the test. The maximum positioning errors is equal to 2.23 m, while the RMS one is 0.88 m. These results demonstrate that our sub-optimal data fusion technique is both efficient and accurate enough to satisfy the large part of non-critical WSNs applications. 

For further evaluation, a classical Kalman filter without fixed gain was used in place of the steady state one. Repeating the analysis, the classical KF outperformed our constant-gain filter with an RMS error of 0.55 m. However, the relatively small difference between the errors shows that the proposed approach, which is more efficient from a computational point of view, can be employed without significantly lowering the localization accuracy.

\section{Conclusions}\label{sec:conclusion}
This article presented a data fusion algorithm suitable to WSNs, taking into accounts data coming from different sources. The approach is based on a steady state KF, where the filter gain is computed off-line. By this way, the procedure complexity is kept low and the filter can be implemented without strong efforts on a typical WSN node. The filter was tuned to fuse radio TOA measurements coming from UWB devices and inertial data sensed from an inertial measurement unit, and its performance has been evaluated in a realistic scenario.
In particular, data about the scenario with variable acceleration ramps have been considered. Firstly, we simulated the scenario replicating the deployment, the radio channel used for ranging and the inertial platform. The results obtained by simulations proved the validity of the novel approach. Further, the fusion process has been applied to the measurements contained in the WPR.B database, collected during a real-world measurement campaign.

Results showed that fusing different sources of information for localization purposes is desirable, since the resulting localization error is greatly reduced. Moreover, our approach is computationally efficient and can be employed in place of more complex filters without significantly affecting  the localization accuracy.

Ongoing works include the filter integration with other positional data sources, such as RSSI readings or maps.
Future work in this area may involve stability issues in presence of longer trajectories or if stopping the target for an extended period of time, when biases and drift effects of the IMU get more relevant. In addition, an analysis of both TOA and IMU estimates errors would be needed to guarantee the stability of the Kalman Filter.

\section*{Acknowledgment}
The authors wish to thank Prof. Dardari, University of Bologna, and the Newcom++
project for providing early access to the N++ WPR.B database.

\newpage


%

  


\end{document}